\documentclass[conference]{IEEEtran}
\ifCLASSINFOpdf
\else
\fi

\usepackage{balance}
\usepackage{graphicx}
\usepackage{color}
\usepackage{colortbl}
\usepackage{xcolor}
\definecolor{dgreen}{rgb}{0.,0.6,0.}
\usepackage{subfig}

\begin{document}
%
\title{On Temporal Regularity in Social Interactions: Predicting Mobile Phone Calls}

\author{
    \IEEEauthorblockN{Mehwish~Nasim\IEEEauthorrefmark{1}, Aimal~Rextin\IEEEauthorrefmark{2}, Numair Khan\IEEEauthorrefmark{3}, Muhammad Muddassir Malik\IEEEauthorrefmark{3}}
    \IEEEauthorblockA{\IEEEauthorrefmark{1}University of Konstanz, Germany
    \\mehwish.nasim@uni-konstanz.de}
    \IEEEauthorblockA{\IEEEauthorrefmark{2}COMSATS Institute of Information Technology, Pakistan
    \\aimal.rextin@comstats.edu.pk}
    \IEEEauthorblockA{\IEEEauthorrefmark{3}National University of Sciences and Technology, Pakistan
        \\nk1394@nyu.edu, muddassir.malik@seecs.edu.pk}
}


%


\maketitle


\begin{abstract}
In this paper we predict outgoing mobile phone calls using a machine learning approach. We analyze to which extent the activity of mobile phone users is predictable. The premise is that mobile phone users exhibit temporal regularity in their interactions with majority of their contacts.  In the sociological context, most social interactions have fairly reliable temporal regularity. If we quantify the extension of this behavior to interactions on mobile phones we expect that caller-callee interaction is not merely a result of randomness, rather it exhibits a temporal pattern. To this end, we tested our approach on an anonymized mobile phone usage dataset collected specifically for analyzing temporal patterns in mobile phone communication. The data consists of $783$~users and more than $12,000$ caller-callee pairs. The results show that users' historic calling patterns can predict future calls with reasonable accuracy.

\end{abstract}


%
\IEEEpeerreviewmaketitle

\section{Introduction and Motivation}

An estimated 261 million Americans own mobile phones. A typical
American makes an average of 5 calls in a single day
\cite{lenhart10}. With a daily average of almost 1.3 billion
communication events and an annual total of 2.45 trillion minutes of
usage \cite{ctia15} in the US alone, mobile phones represent one of the most
commonly used communication medium. Judging from the trend in recent
years, the user base of cell phones can be expected to further
increase in the future. Hence, a deeper understanding of the temporal
structure of mobile phone communication would allow us to optimize and
streamline a technology that has penetrated the very fabric of human
life.

Users generally make phone calls in two ways: either by selecting the
callee from a contact list, or through the call log \cite{bergman2012you},\cite{calladdressbook},\cite{addressbookbad}. The former
displays contacts in alphabetical order with no consideration of past
calling behavior. While most mobile phones offer the capability of
selecting certain contacts as \textit{favorites}, the favorites list
is, however, still a static list, requiring active intervention by the
user in order to update. Call logs, on the other hand, do take past
user behavior into account, displaying called numbers in reverse
chronological order. The model of user behavior assumed by call logs
is, nonetheless, highly simplistic. It supposes that the likelihood of
calling a particular contact $c$, is $P(c)$, which is a monotonically decreasing
function of the time elapsed since last contact. Sociologists have,
however, shown that human life is temporally organized and that most
social interactions have fairly reliable temporal regularity
\cite{zeru85}. This implies that $P(c)$ could be periodic. Such an implication, if correct, would allow for the design of a
considerably more efficient calling interface than what is provided by
either contact lists, or chronological call logs. From the service
providers' perspective, it would enable them to predict users'
behavior, thereby allowing targeted and personalized product
recommendations. This would, in turn, lead to greater customer loyalty
\cite{Had14}. In addition, the ability to predict periods of high
usage would lead to better load balancing and, hence, better service
quality.

Based on the assumption that one could predict the users' calling behavior using temporal features, we use a machine learning approach that aims at predicting the future calls made by a user, based on his/her past calling behavior. 

\textbf{Contributions}
Our contributions are as follows:
\begin{enumerate}
\item We focus on predicting outgoing calls by modeling the problem as a multiclass classification problem. Our approach can generate real time predictions for the outgoing calls based on the historical calling patterns. 
\item Sociological theories show that most
social interactions have fairly reliable temporal regularity. Based on this premise we have identified features that can accurately predict the future calls of a user.
\item There are about $150$Mn. mobile phone users in Pakistan and $31\%$ of them have smartphones. Till date, we did not find any substantial analysis on the data of this huge user base. We test our approach on a large dataset which we specifically collected for analyzing calling behaviour of mobile phone users based in Pakistan. 

\end{enumerate}
 
\textbf{Paper organization}
The remainder of the paper is organized as follows: In section \ref{sec:related}, we briefly summarize the existing work on finding periodic patterns in human social interaction and previous call prediction methodologies. In section \ref{sec:collection}, we
discuss a smartphone application (app) that was used for collecting data and report the data statistics. This is followed by section \ref{sec:methodology} that overviews our methodology. In section \ref{sec:analysis} we present our results. In section \ref{sec:discussion} we discuss our findings and compare our results with previous approaches. This is followed by a conclusion and future work section.

\begin{figure}[t]
 \centering
\includegraphics[width=0.35\textwidth]{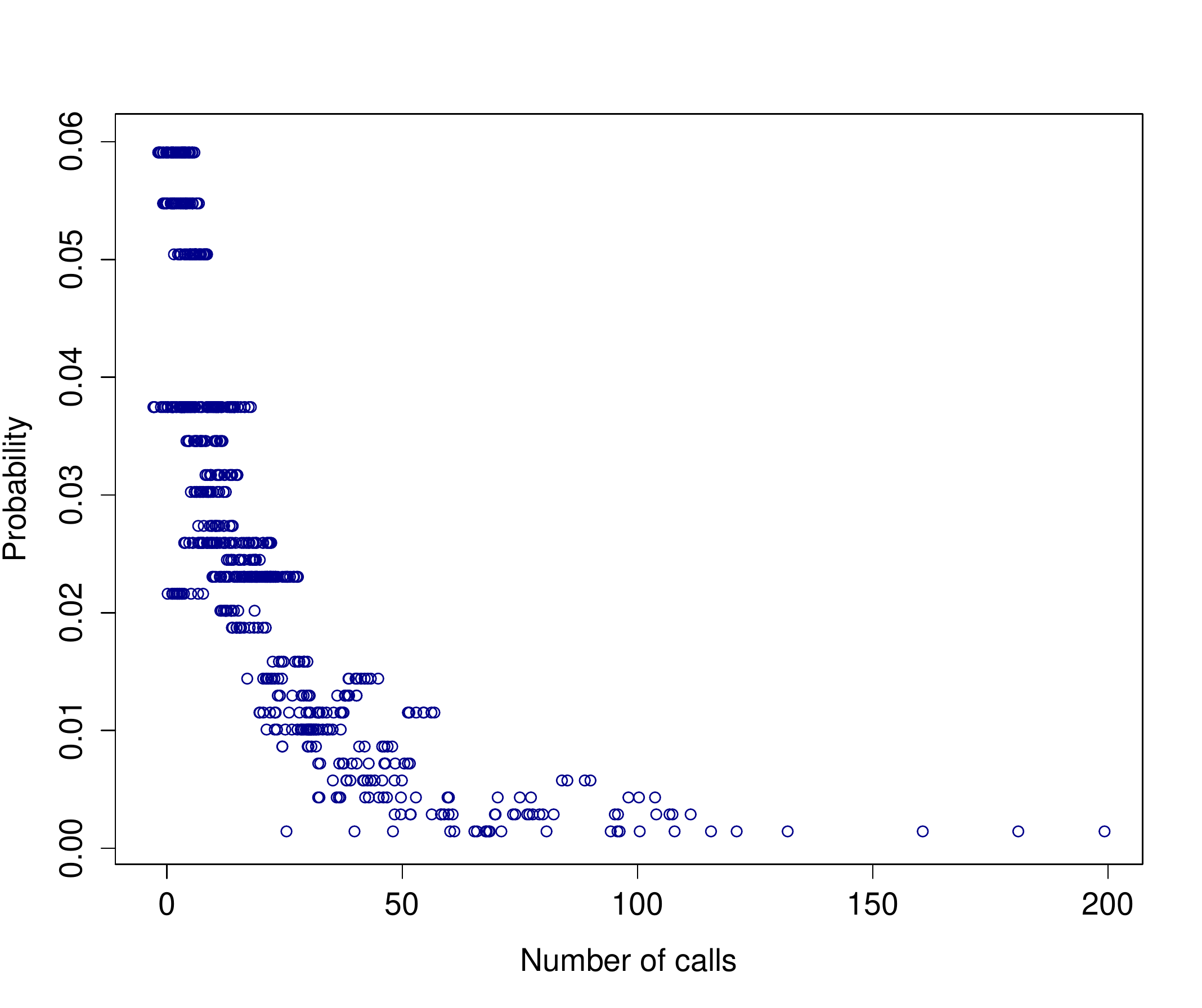}
\caption{Average number of calls per day.}
\label{fig:dailycalls}
\end{figure}

\section{Related Work}
\label{sec:related}
Call log data can provide insights into the underlying relational dynamics of
societies, evolution of relationships over time and, can also help in prediction of social network structures \cite{inferfriendship}. Data of calling patterns has been
used to infer friendships relations and uncover individual and
collective human dynamics \cite{inferfriendship, candia2008uncovering,
  jiang2013calling}. Call-volume data has been used to explore whether
the distribution of calls in an urban population follow routine
patterns or not, and whether the variation of such patterns in
different parts of the city can be explained \cite{urbanmobility}. While studying social network turnover, Aledavood et
al. \cite{aledavood15a},\cite{aledavood15b} found that individual calling
and messaging behavior follows a circadian rhythm. Their study of 24
subjects revealed that the frequency and entropy of communication
displays a distinct daily pattern that remains persistent over
time.

Several call prediction models have been proposed in the literature. Phithakkitnukoon et al. \cite{phithakkitnukoon2011behavior} predicted the outgoing and incoming calls on Reality Mining dataset \cite{eagle06} based on most recent calling data. Out of the $94$ datasets, they used a small subset of $30$ users for performance evaluation. Haddad et al.
\cite{Had14} presented a probabilistic model that uses call frequency to
predict incoming and outgoing calls for each individual
contact. Their underlying assumption is that the calling
behavior of users can indeed be modeled as a periodic phenomenon. They have modeled the users' recurrent phone call behavior using Poisson process.  The
authors tested their model on a large sample by making it available as
a mobile application. Several recent studies of human behavior indicate that the timing of
communication events is characterized by long dormant periods
interspersed with bursts of high activity \cite{barabasi05, jo12, wu10}. Barabasi \cite{barabasi05} attributes this bursty non-Poisson
character of human behavior to a priority-based queuing process. This
view is supported by Jo et al. \cite{jo12} who show that burstiness
remains in mobile communication data even after circadian and weekly
patterns have been removed, precluding the attribution of periods of
inactivity to nights or weekends. They conclude that burstiness
results from non-homogeneity in human task execution mechanisms. Another study conducted by Kim et al. \cite{kim2013analyzing} on a large dataset from North-American users also suggests that the caller-callee behavior cannot solely be modeled using the Poisson distribution. Kim et al. \cite{kim2013analyzing} performed a comprehensive
analysis of interaction frequencies by analyzing the communication activity of
over one million pairs of mobile phone subscribers from a American
cellular service provider. They focus on studying the impact of family relationship on communication patterns. Based on frequency of information exchange, they classified the user-pairs into three classes characterized by the inter-arrival times between calls made between pairs. 


\begin{figure}[t]
 \centering
\includegraphics[width=0.35\textwidth]{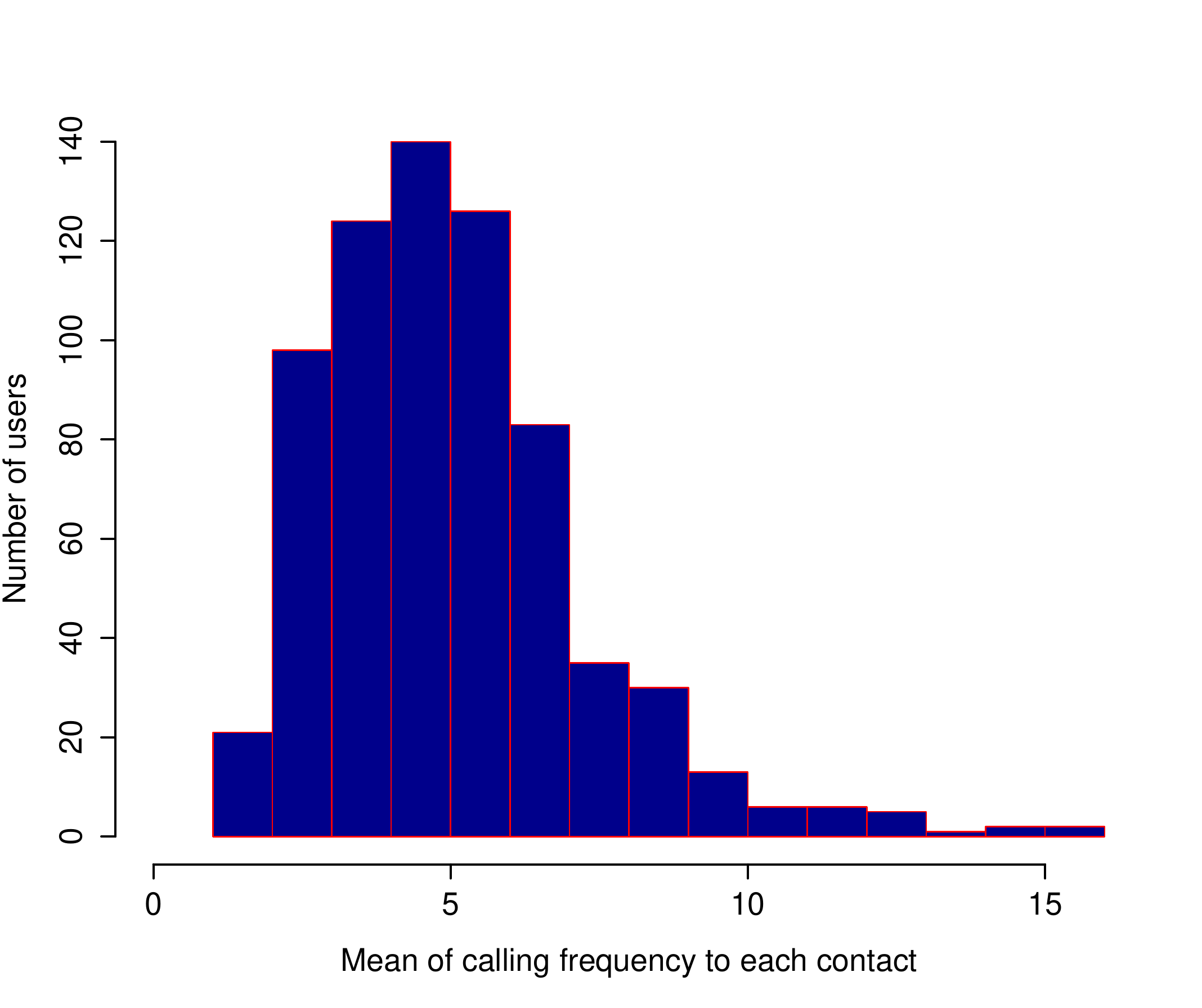}
\caption{Mean of average number of calls per user per contact.}
\label{fig:meanfr}
\end{figure}

\section{Data}
\label{sec:collection}

We tested our approach on the following two datasets. 

\subsection{Reality Mining Dataset}The first data set is from Reality Mining experiment \cite{eagle06}, referred to as: `Reality Mining dataset'. This dataset contains data from $94$ users which were either students, faculty or staff at the Massachusetts Institute of Technology, USA. 

\subsection{Smartphone Dataset}
We collected the second dataset using a smartphone app referred to as: `Smartphone dataset'. Data collection was limited
demographically to users of smartphones running the Android OS, and
geographically to the country of Pakistan. While industry sources estimate that Android users represent 68\% of the total smartphone population in
that country, extensive market surveys are lacking and, hence,
conclusive judgments about the qualitative nature of the sample
cannot be made.

To make its value proposition more attractive, the app presented users
with the most economical mobile subscription service for their needs based on past
calling behavior. These subscription services - also referred to as
``packages'' in the local parlance - differ primarily in the calling
rates they offer during specific hours of the week. A recommender
system for similar telecom products was developed by Zhang \textit{et
  al.} \cite{zhang13}. But, where they used fuzzy-set techniques to
select the most economical product, our recommendations are based on a
simple simulation run with the users' call history. Including this
additional functionality in our data-gathering app not only expanded
our sample set, but we also expected it to mitigate the volunteer bias
natural in such survey data collection methods. Users were notified that their call data would be used for academic
research purpose. 

\textbf{Data Statistics}
The process of Smartphone dataset collection lasted from July 28, 2015 till September 24, 2015. The data was collected from April 19, 2015 till September 23, 2015.  The data consists of $783$~users (egos) with $229,450$~communication events. The data for each ego was grouped according to the contact the communication event was initiated to - an \textit{alter} - thereby, yielding a over 12,000 ego-alter\footnote{Ego is the focal actor who installed the application. Alters are his/her contacts.}pairs in the smartphone dataset, and over 2000 pairs in the Reality Mining dataset. The probability distribution function (PDF) of number of calls per user are shown in Figure \ref{fig:dailycalls}. On average, each user made or received $\approx 22$ calls per day.

We conducted a pretest on the aforementioned mobile phone datasets (Smartphone dataset and Reality Mining dataset) for determining whether caller-callee pairs have a regular calling pattern. The communication between mobile phone users  was modeled as a time series data analysis problem. In many time series, it is plausible to expect that the $K$ recent data points are likely to have an influence on the future data points. In order to identify whether the ego-alters communication data has a pattern, autocorrelation was used which is a type of correlation statistic specifically for correlating the recent data point to other data points in the series.
For each ego-alter pair, an hourly and a daily autocorrelation measure
was calculated using the Ljung–Box test (Q test) where a $p-value < 0.05$ means there is autocorrelation.
 The Ljung–Box test, also known as a \textit{portmanteau} test, is a
function of the accumulated sample autocorrelations $r_k$, up to any
specified time lag $m$ \cite{ljung1978measure} (we checked the autocorrelation up to six lags. If the lag is too small, the Q-test may not detect serial correlation at high-orders. If its too large, the test may have low power since the significant correlation at one lag may be diluted by insignificant correlations at other lags.).  
 As the hourly
autocorrelation measure may have been biased by lack of activity at
night hours, a third autocorrelation for communication events between
7am and 8pm was also calculated. Results obtained on these datasets quantitatively show that a reasonable number of ego-alter pairs exhibit autocorrelation. More than~50\% of ego-alter pairs in the Reality Mining dataset exhibit a daily as well as hourly periodic calling behaviour. In the Smartphone dataset,  few ego-alter pairs exhibited daily temporal autocorrelation, but it was observed that there was indeed an indication of periodic calling at finer levels (hourly basis). The difference in results obtained from the datasets could be an artifact arising from the shift in communication from mobile phone calls/text messages to smartphone instant messengers in recent years. Reality Mining dataset was collected when other means of smartphone communication such as Whatsapp, Viber, Facetime, etc. did not exist. Another tenable explanation could be the bias in the datasets. Contrary to the Reality Mining dataset that contains data from students or faculty of MIT media lab with fixed university schedule, the smartphone dataset contains data from general population with different demographics.

Findings in \cite{ourpaper} suggest that mobile phone users exhibit periodicity in their calling behavior (at daily or hourly levels). Based on these findings, we set the focus of this paper on predicting the future outgoing calls using time based features.

\begin{table}[b]
\caption{Proportion of correctly predicted calls.}
\centering
\begin{tabular}{ l | c | c | c| c}
\hline

        Dataset       & $ \epsilon(15Min)$ & $\epsilon(1Hr)$ & $\epsilon(10Hr)$ & $\epsilon(24Hr)$\\
\hline
\hline
Reality Mining & 0.40 &  0.44   & 0.60 & 0.70 \\
\hline
Smartphone     & 0.26 & 0.31 & 0.53 & 0.63 \\
\hline
\end{tabular}

\label{tab:predictionaccuracy}
\end{table}

\section{Methodology}
\label{sec:methodology}
We model the problem as a supervised multi-class classification problem. \textit{Multiclass}, also known as multinomial classification is the problem of classifying instances into one of the more than two classes. In every ego profile, alters that have been called in the past are the classes. The classifier predicts at a given time and day which number(s) a user is likely to call. Each dataset and further each ego profile is evaluated independently. 

\subsection{User Behavior Model}

Let $U$ be the set of all egos in the dataset. For every $u_i \in U$, let $X_i$ be the set of contacts.
Formally, we define the call prediction problem as follows: Given the historical communication events, $\{Y_i(t_0), ..., Y_i(t_{(n-1})\}$ consisting of outgoing, incoming and missed calls that occurred at time $t_0, ..., t_{n-1}$, for a user $u_i \in U$, predict which contact$ \{x_1, ..., x_m\} \in X_i$,  $u_i$ is going to call at time $t_n$. 

As a first step, for every ego we find the outgoing calls distribution. There is a general observation that there are fewer contacts who are called more often and a lot many contacts who are called less often. We observed that the outgoing calls are not normally distributed.  We then test the null hypothesis that the outgoing calls distribution for each ego is \textit{exponential}. We use the Kolmogorov-Smirnov test (KS test), which is a nonparametric test of the equality of continuous, one-dimensional probability distributions. It can be used to compare a sample with a reference probability distribution. We found that KS test fails to reject the null hypothesis at the 5\% significance level for $80.08\%$ of Reality Mining users and for $87.64\%$ of Smartphone users. Thus, for each ego, we remove the callees who are called less than a certain threshold. For our experiments we selected the mean of the calling frequency as the cut-off threshold (Figure \ref{fig:meanfr} shows a histogram of the mean of calling frequency per user per contact). It is plausible to remove those callees which are very sporadically contacted. Mobile phone users have various kinds of contacts in the contacts list such as friends, acquaintances, family members, workplace contacts, services related contacts, etc \cite{bergman2012you}. Some of these contacts are frequently called, others are occasionally called and some contacts are not called in a long time. This view is also supported by \cite{kim2013analyzing}. This last category of contacts, possibly does not exhibit temporal regularity at the time scale at which we are studying the problem. Hence, it is reasonable to remove them by setting a threshold on the communication frequency. In the final analysis $10,383$ caller-callee pairs are analysed in the Smartphone dataset and $1851$ pairs are analyzed in the Reality Mining dataset.

\subsection{Features }
We extracted different temporal features which we believe are most accurate in predicting the outgoing calls based on the premise that majority of the caller-callee pairs either exhibit daily or hourly temporal patterns in their interaction. For every call, the explanatory variables(features) are :time of the day (correct to the nearest minute), weekday (Sunday-Saturday), Night-call (true or false), last dialed numbers, timestamp of last dialed numbers and direction of the call (incoming, outgoing or missed). The class label is the called party's phone number.



\subsection{Classification}

We classified our data using the Support Vector Machines (SVM) classifier, using the implementation
available in R \cite{rsvm}. We divided the data (each ego's call log) into training and test sets. We use $80\%$ of the data for training the model and the remaining $20\%$ data for predicting future calls\footnote{The Smartphone dataset contains data from April till September 2015. On average, the training set contained data approximately from April-August 2015 (about 16 weeks), used to predict calls made between August-September 2015 (about 5 weeks). On average, training the model took 48.35 milliseconds for each ego, on a Lenovo X1 Carbon Notebook with Intel Core i-7 CPU(2GHz) and 8GB of RAM.}. We have ensured mutual exclusivity between training and test data. We use a linear combination of the calling features for prediction. For every call in the test set, the classifier outputs probabilities against each class.

\subsection{Evaluation Metric}

We have used the following two evaluation criteria for performance evaluation: 

\begin{enumerate}
\item In a hypothetical situation, whenever a user presses the call button (or opens the calling interface), at time $t_n$, a list of contacts is displayed. Our classifier outputs probabilities for each class (contact) a user is likely to call at time $t_n$. These probabilities are computed and sorted and a list of contact numbers with the highest probabilities is displayed. We call this list, `top-$k$ recommendations'. We have also compared the performance of our approach with top-$k$ most frequently called numbers and with last-$k$ calls. We calculate the probability that $u_i$ is going to call $x_j \in X_i$, given that $\theta_j$ amount of time has elapsed since the last communication. We denote this probability by $P(x_j | \theta_j)$. We observed that when $\theta_j$ is small or when the last communication event was a missed call from $x_j$, the probability to call $x_j$ is high. For performance comparison, we generate a final list of most likely numbers to be dialed at any given time (within the next hour) based on the results produced by the classifier.  
 
\item In the second evaluation method, we measure the proportion of calls that are predicted within a certain error threshold($\epsilon$). For a given time, a `single phone number' is predicted which the user is likely to call. We then measure how well the number is predicted with regards to different time-deviation thresholds. 
\end{enumerate}

\begin{figure*}[t]
 \centering
\subfloat[]{\includegraphics[width=4cm]{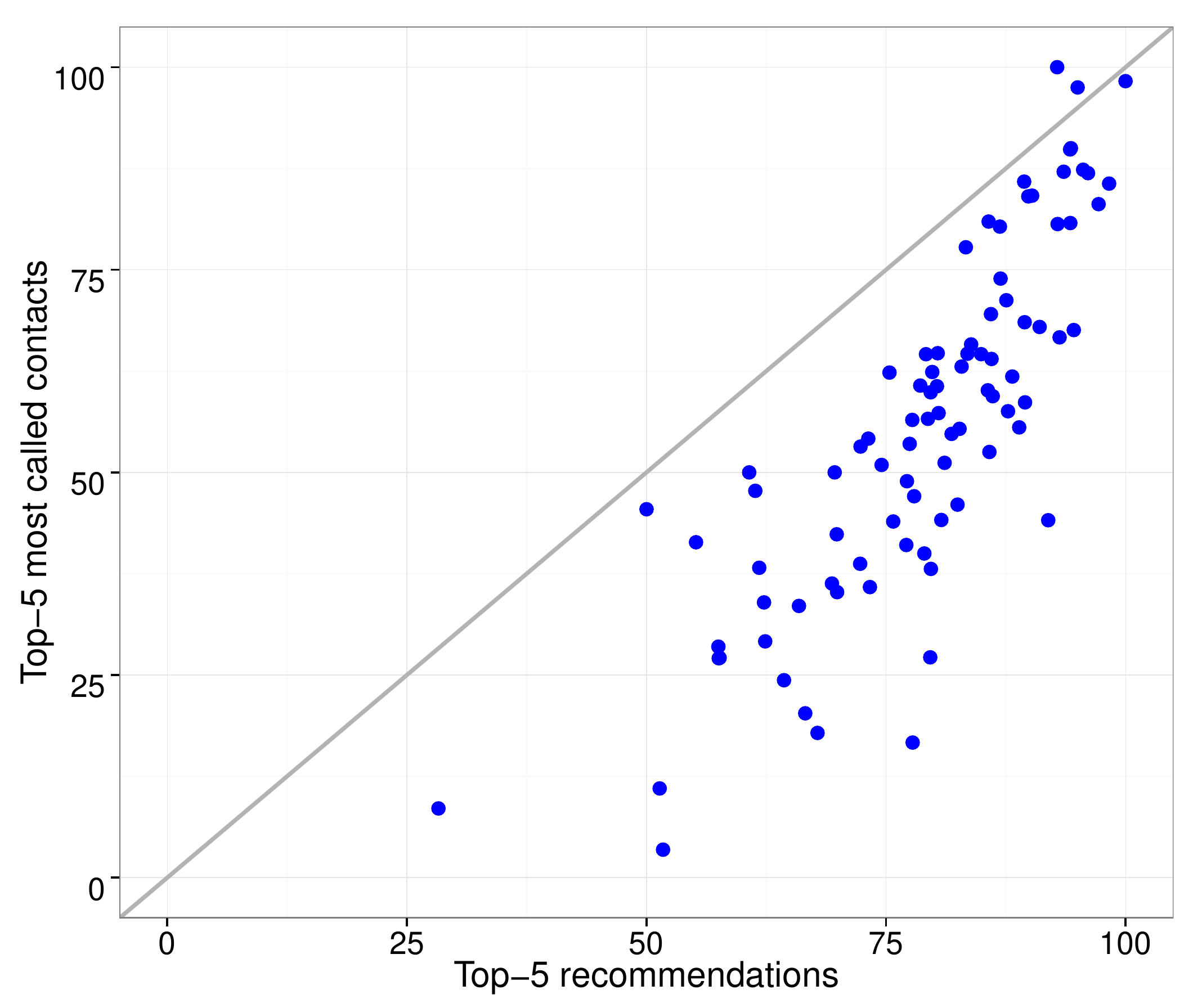}\label{subfig:top5rm1}}\hspace{1em}
\subfloat[]{\includegraphics[width=4cm]{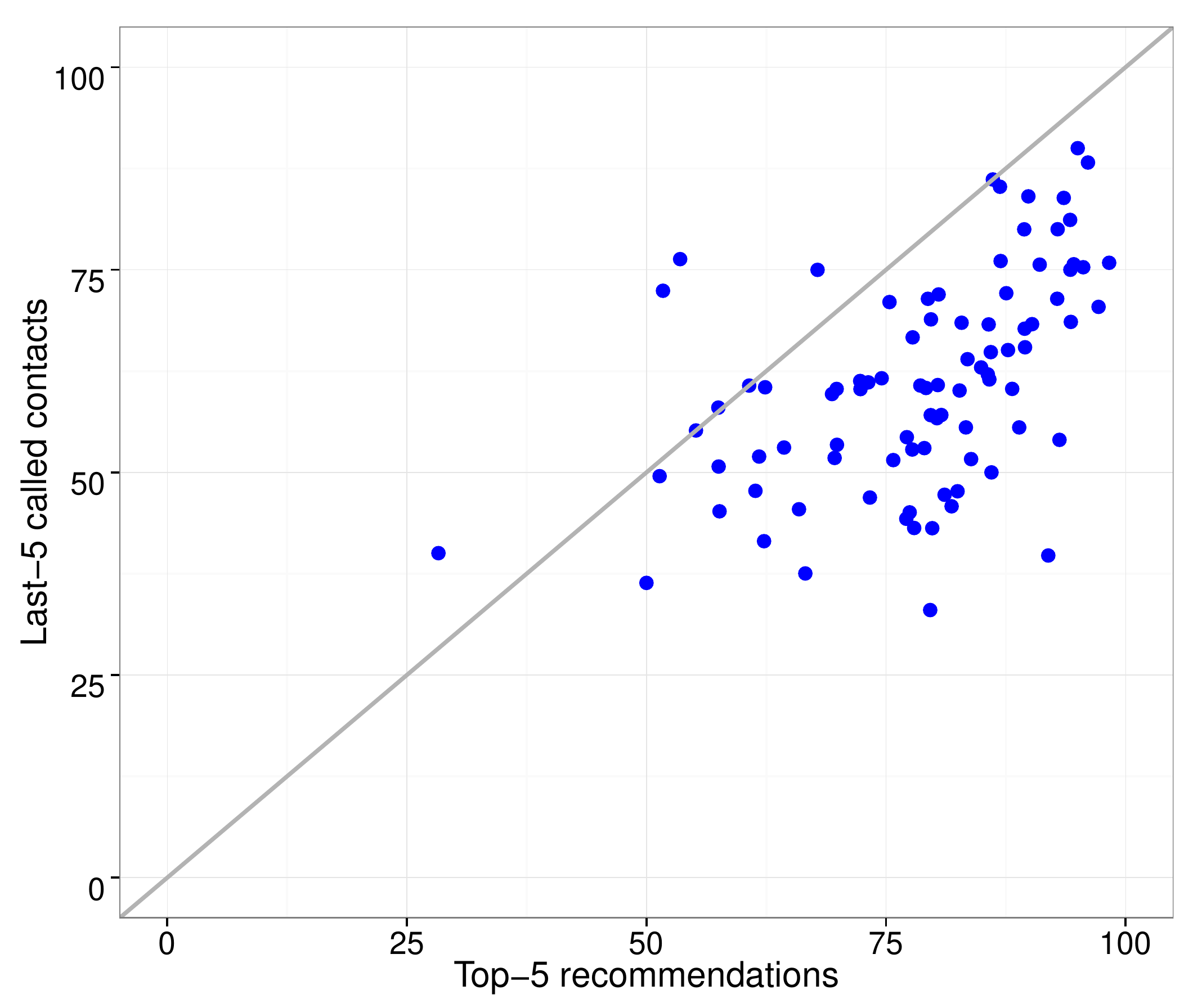}\label{subfig:top5rm2}}\hspace{1em}
\subfloat[]{\includegraphics[width=4cm]{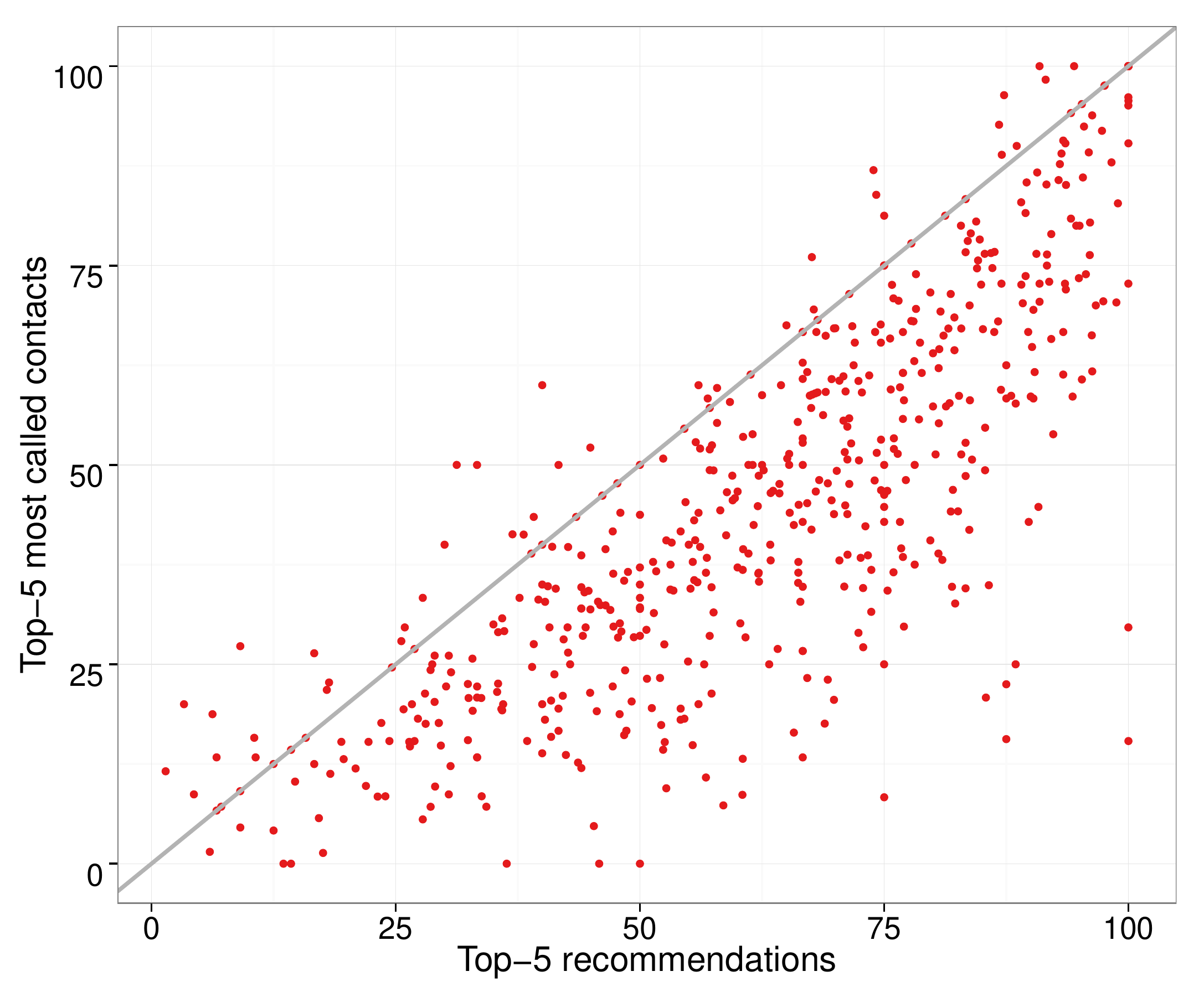}\label{subfig:top5sp1}} \hspace{1em}
\subfloat[]{\includegraphics[width=4cm]{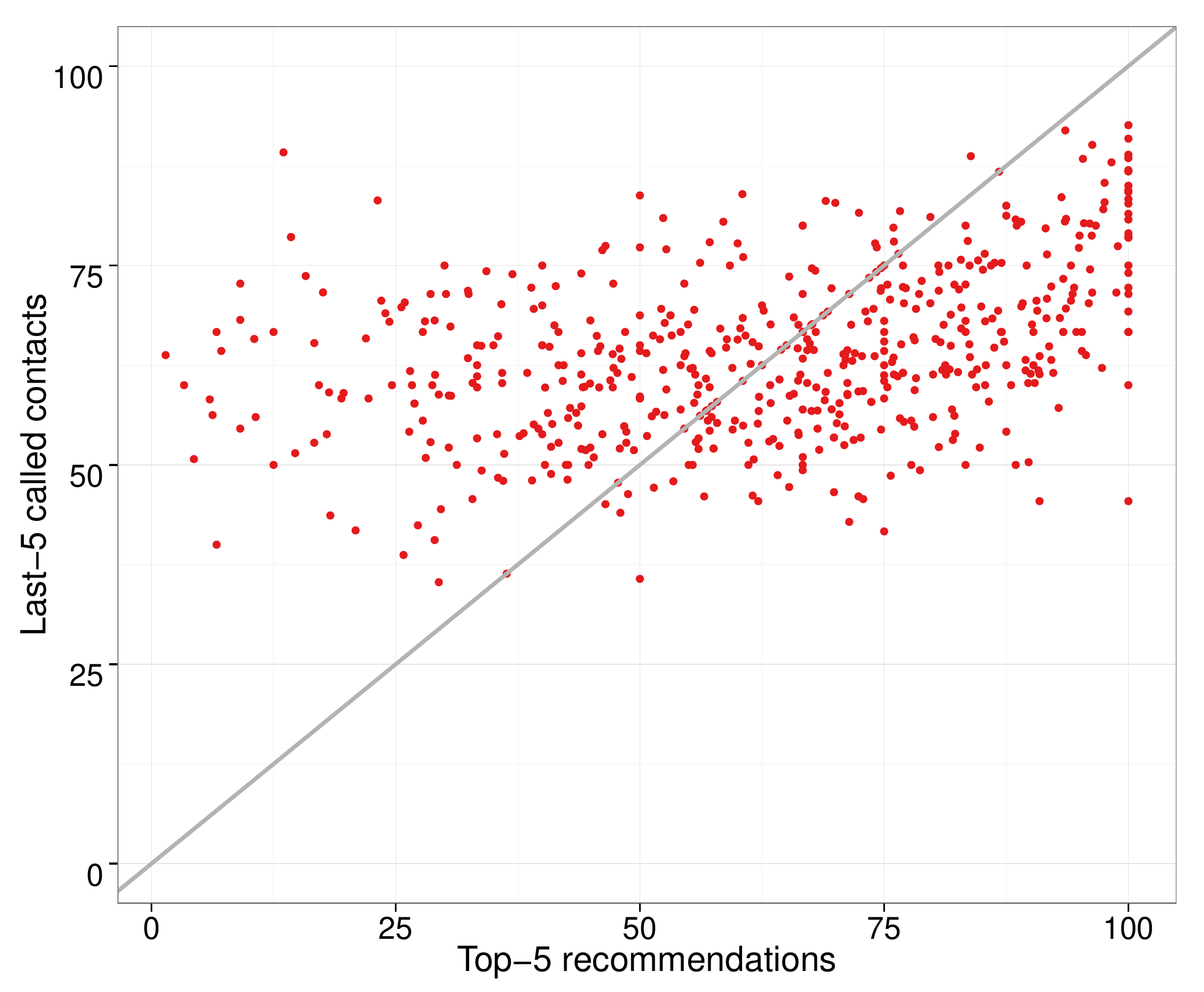}\label{subfig:top5sp2}}\\
\subfloat[]{\includegraphics[width=4cm]{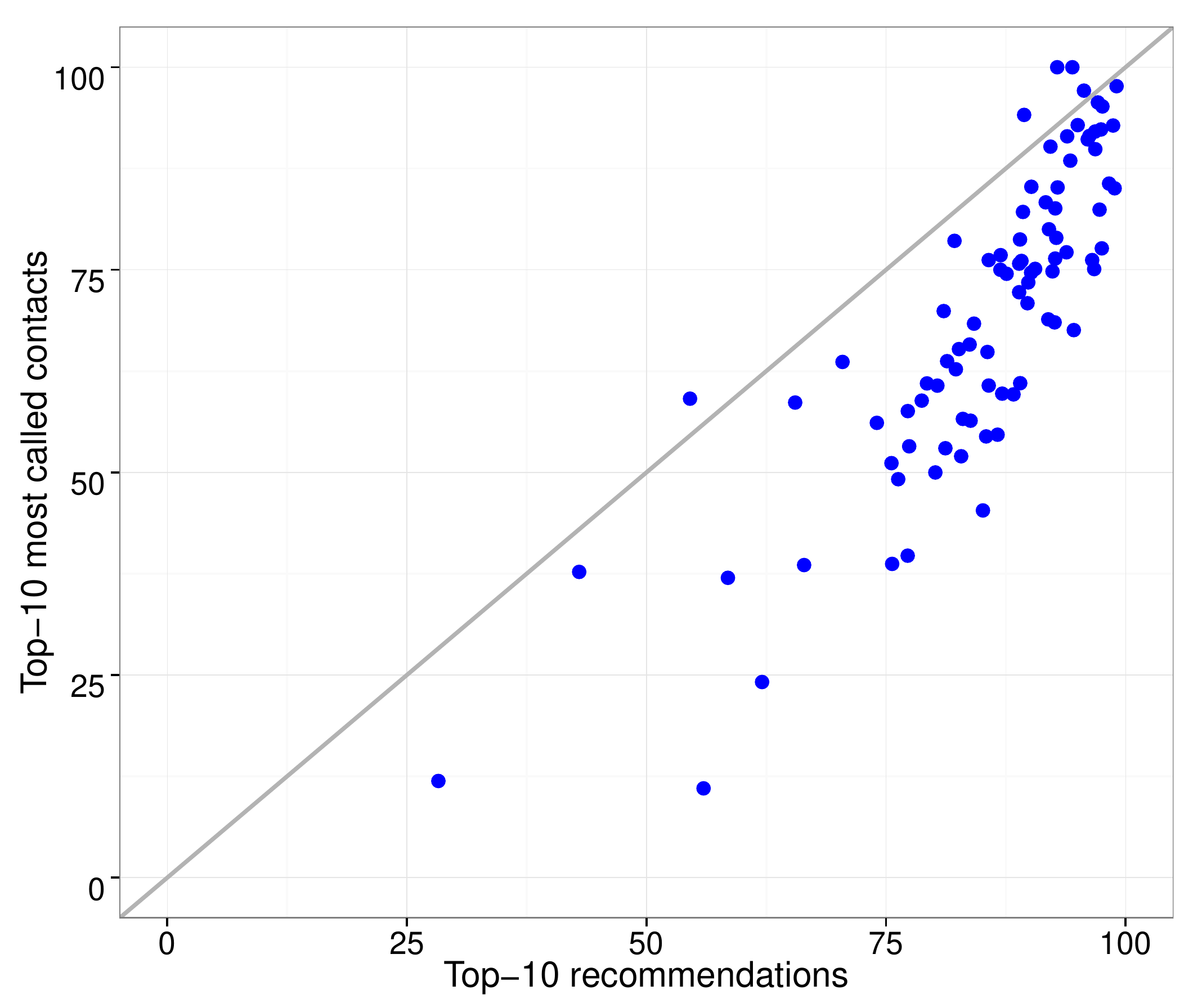}\label{subfig:top10rm1}}\hspace{1em}
\subfloat[]{\includegraphics[width=4cm]{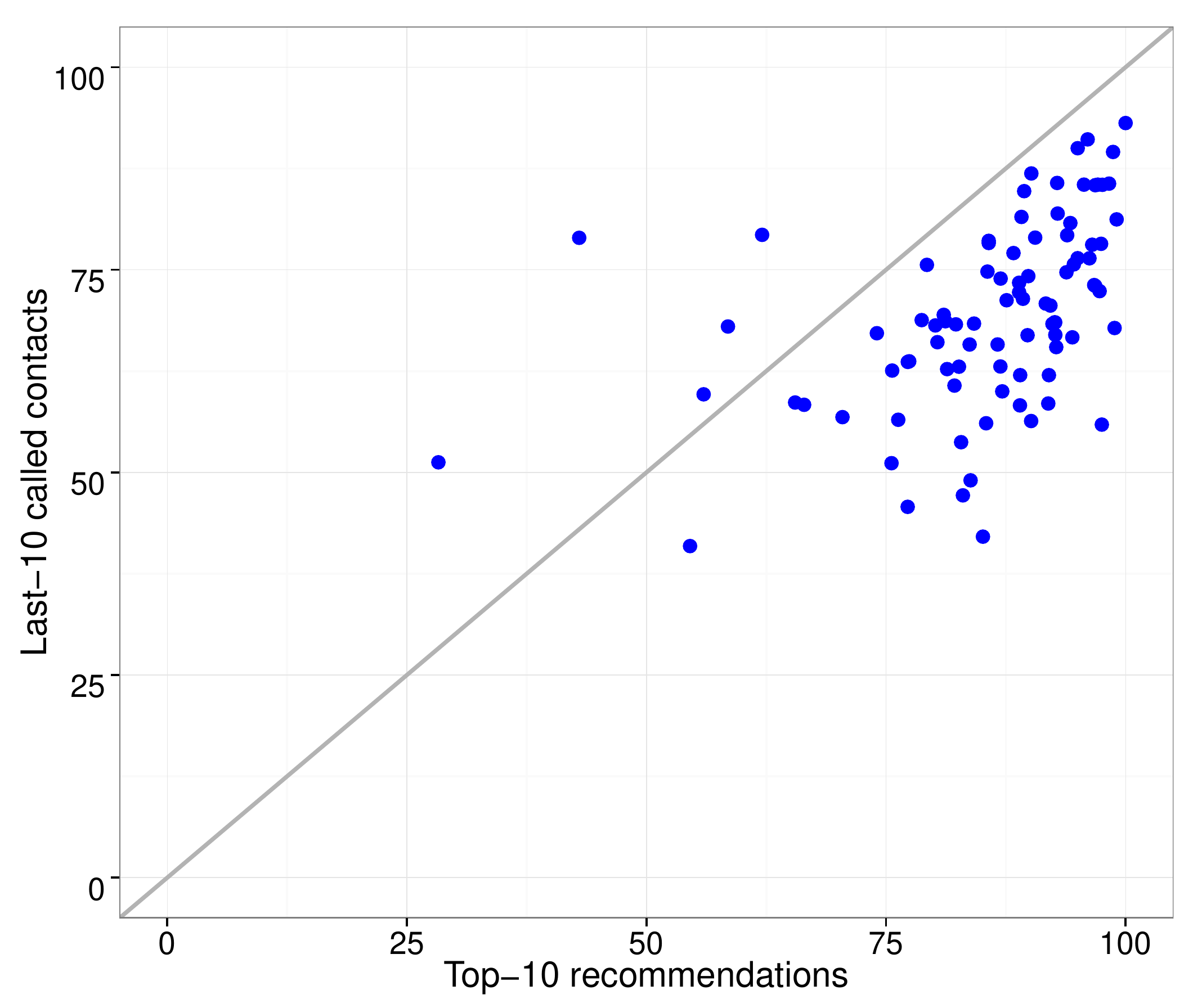}\label{subfig:top10rm2}}\hspace{1em}
\subfloat[]{\includegraphics[width=4cm]{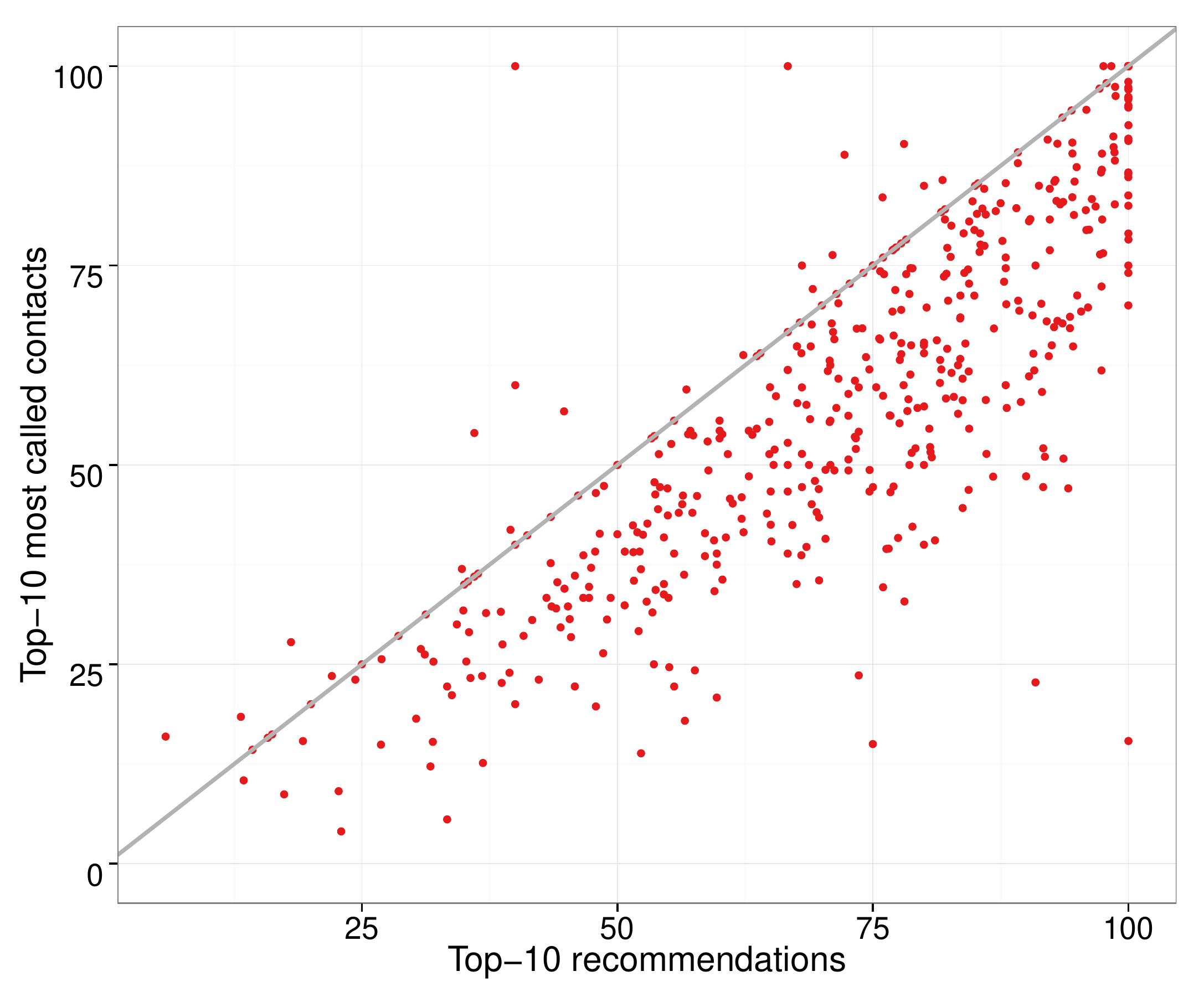}\label{subfig:top10sp1}}\hspace{1em}
\subfloat[]{\includegraphics[width=4cm]{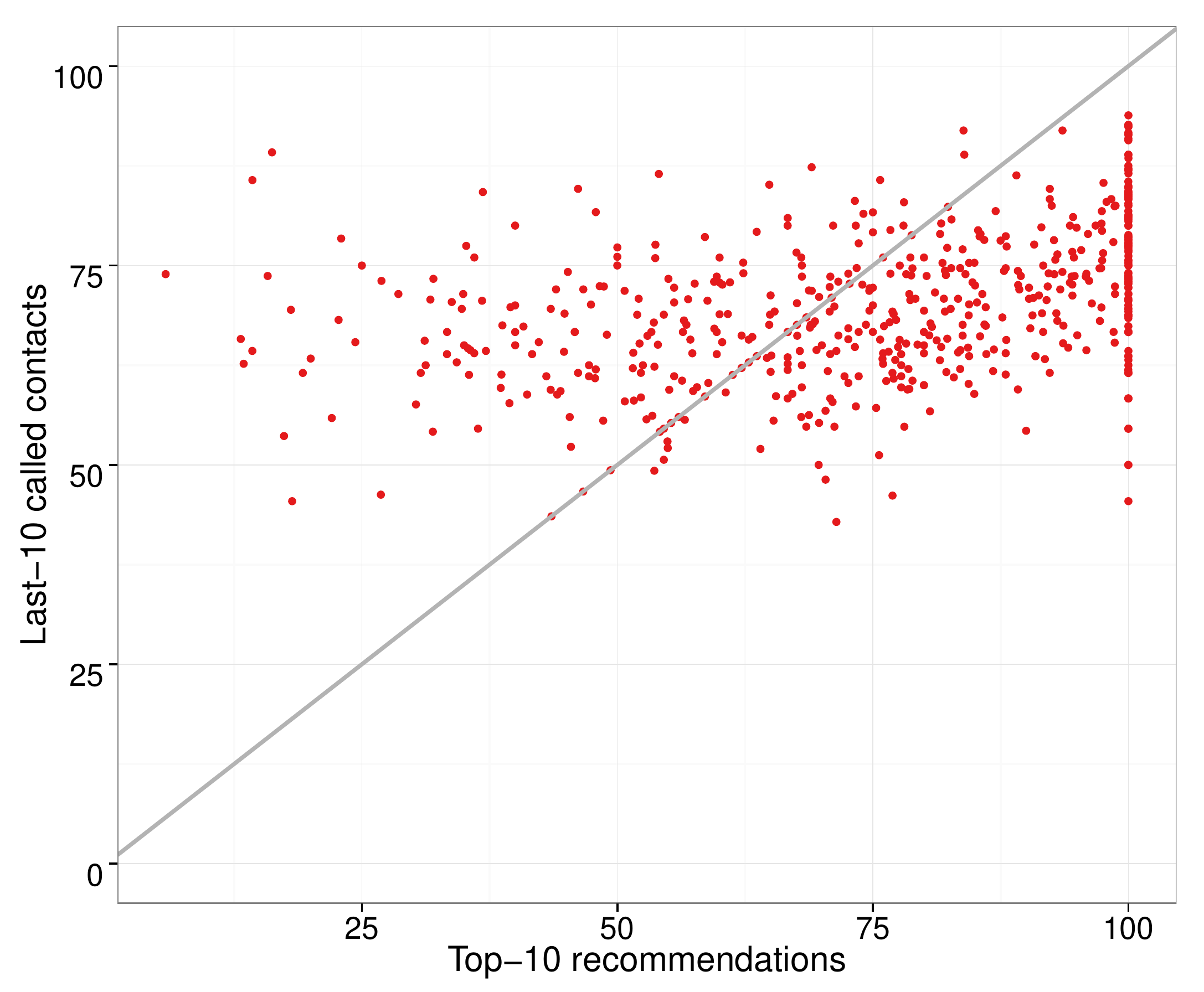}\label{subfig:top10sp2}}\\
\subfloat[]{\includegraphics[width=4cm]{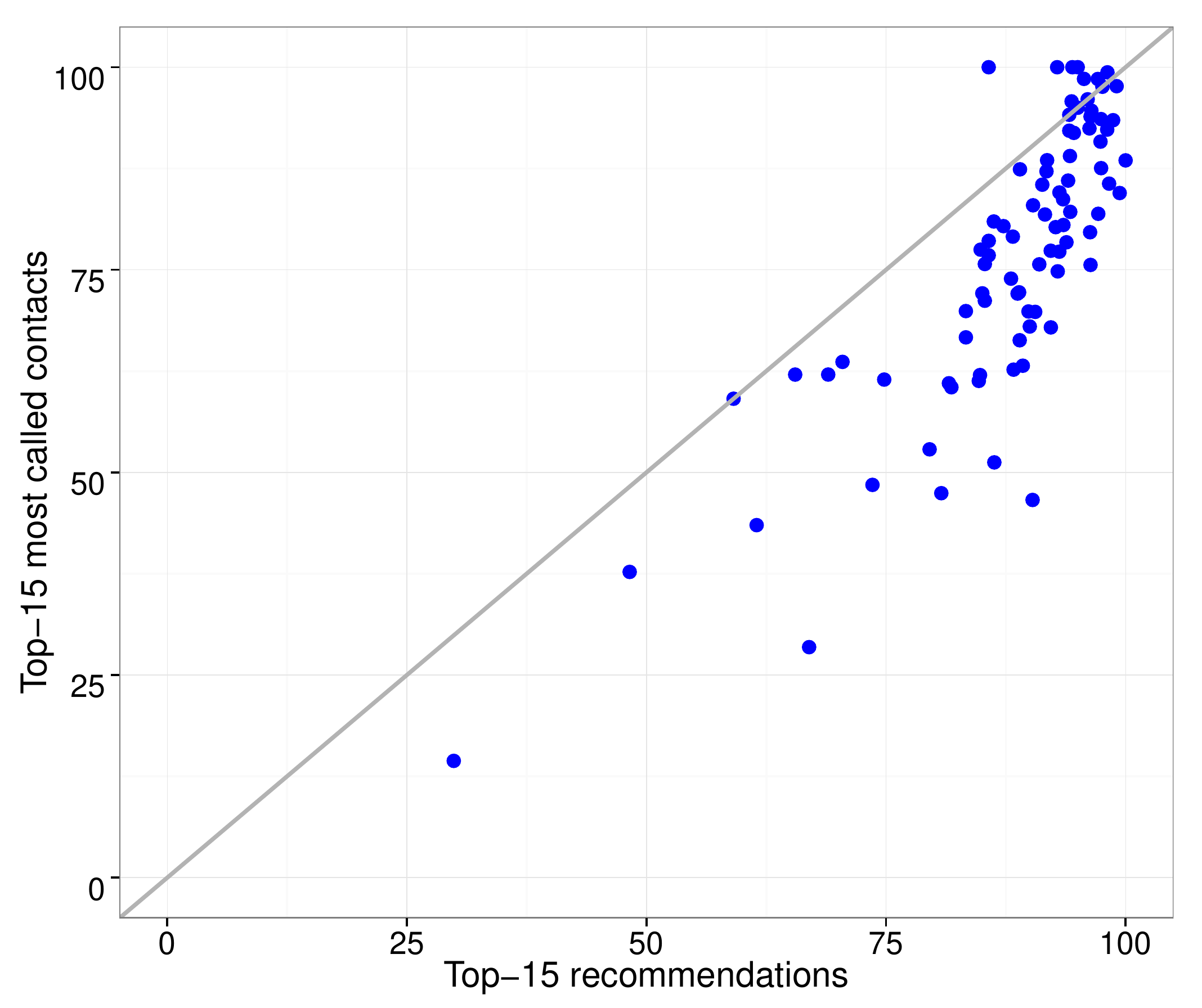}\label{subfig:top15rm1}}\hspace{1em}
\subfloat[]{\includegraphics[width=4cm]{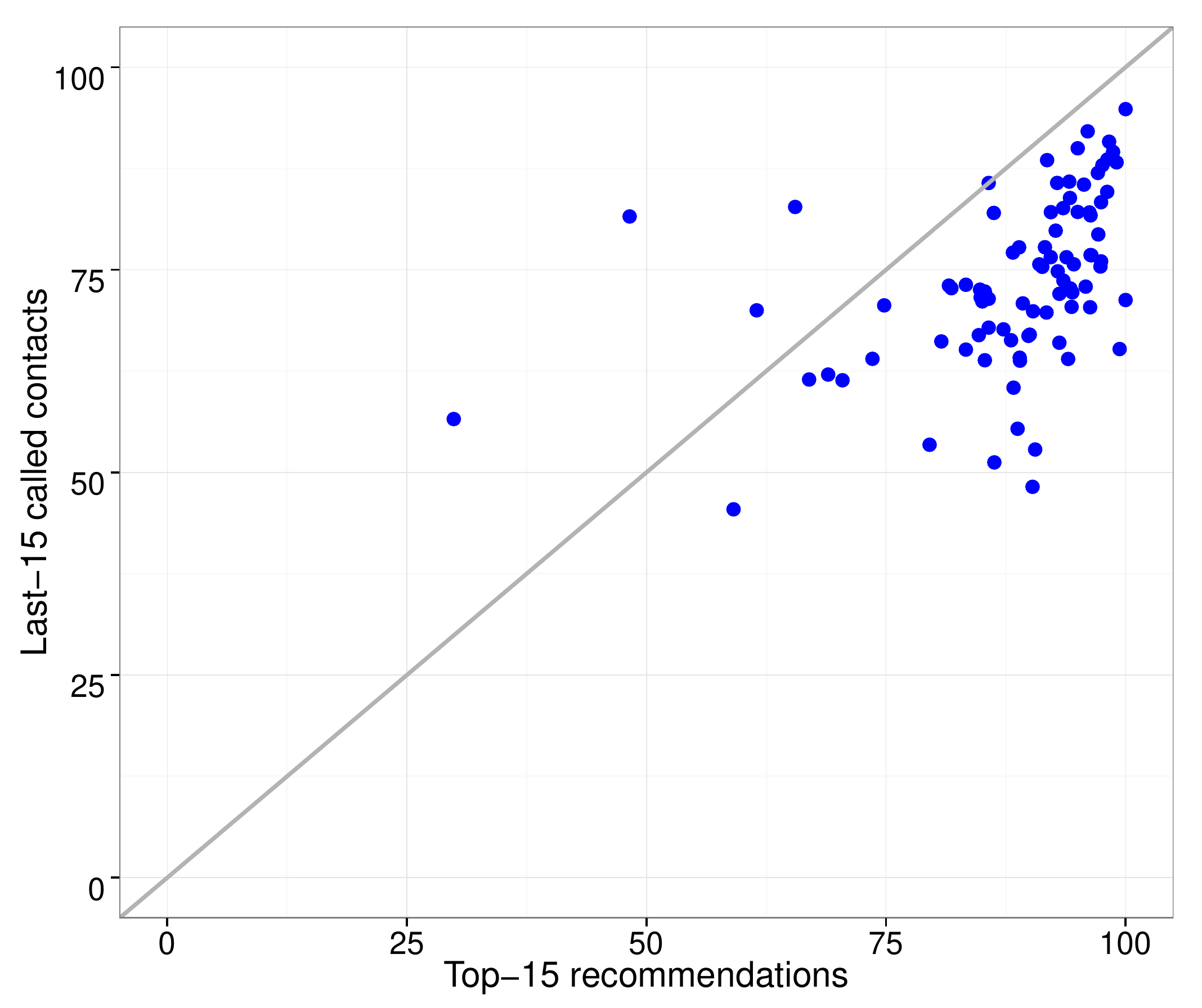}\label{subfig:top15rm2}}\hspace{1em}
\subfloat[]{\includegraphics[width=4cm]{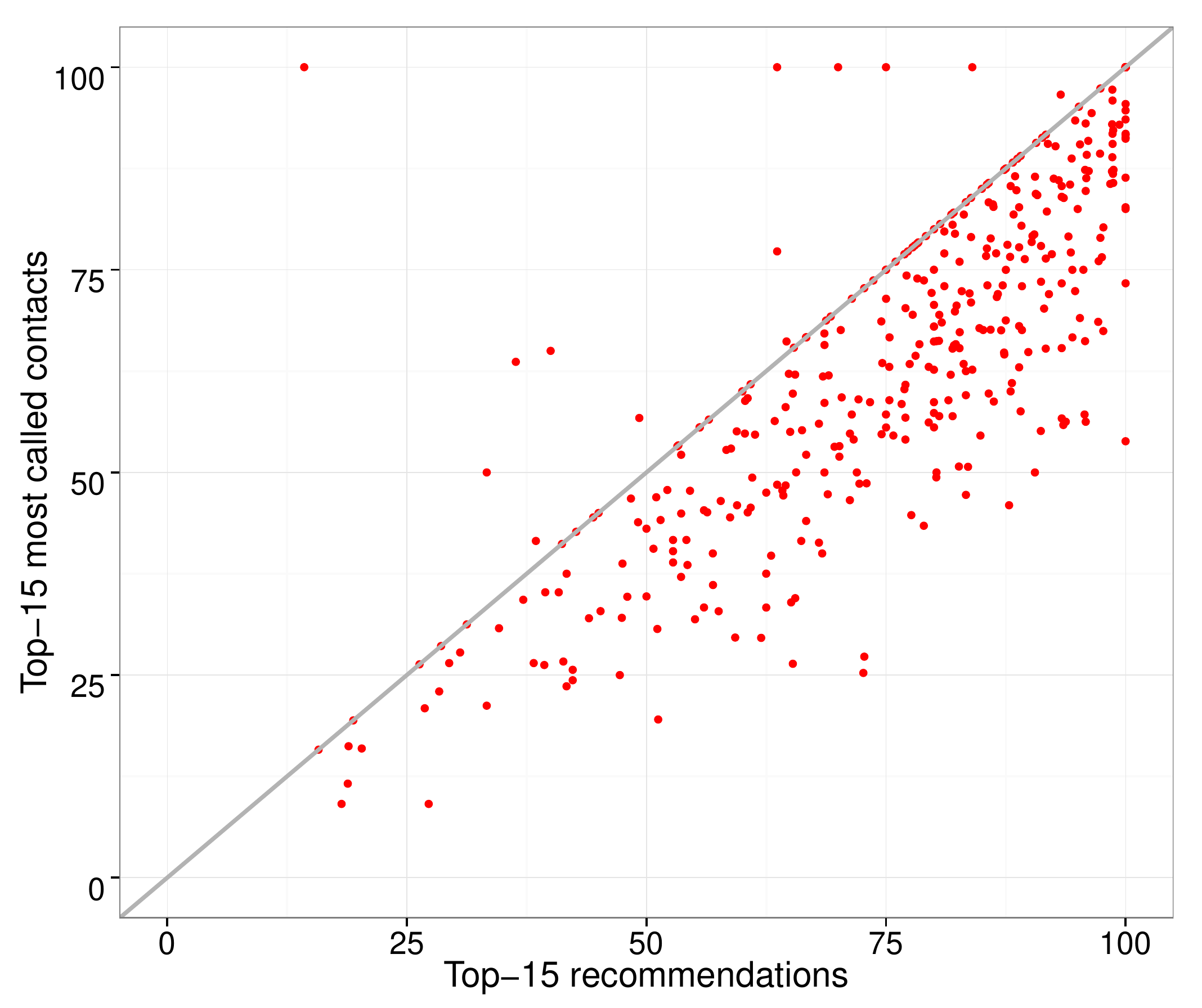}\label{subfig:top15sp1}}\hspace{1em}
\subfloat[]{\includegraphics[width=4cm]{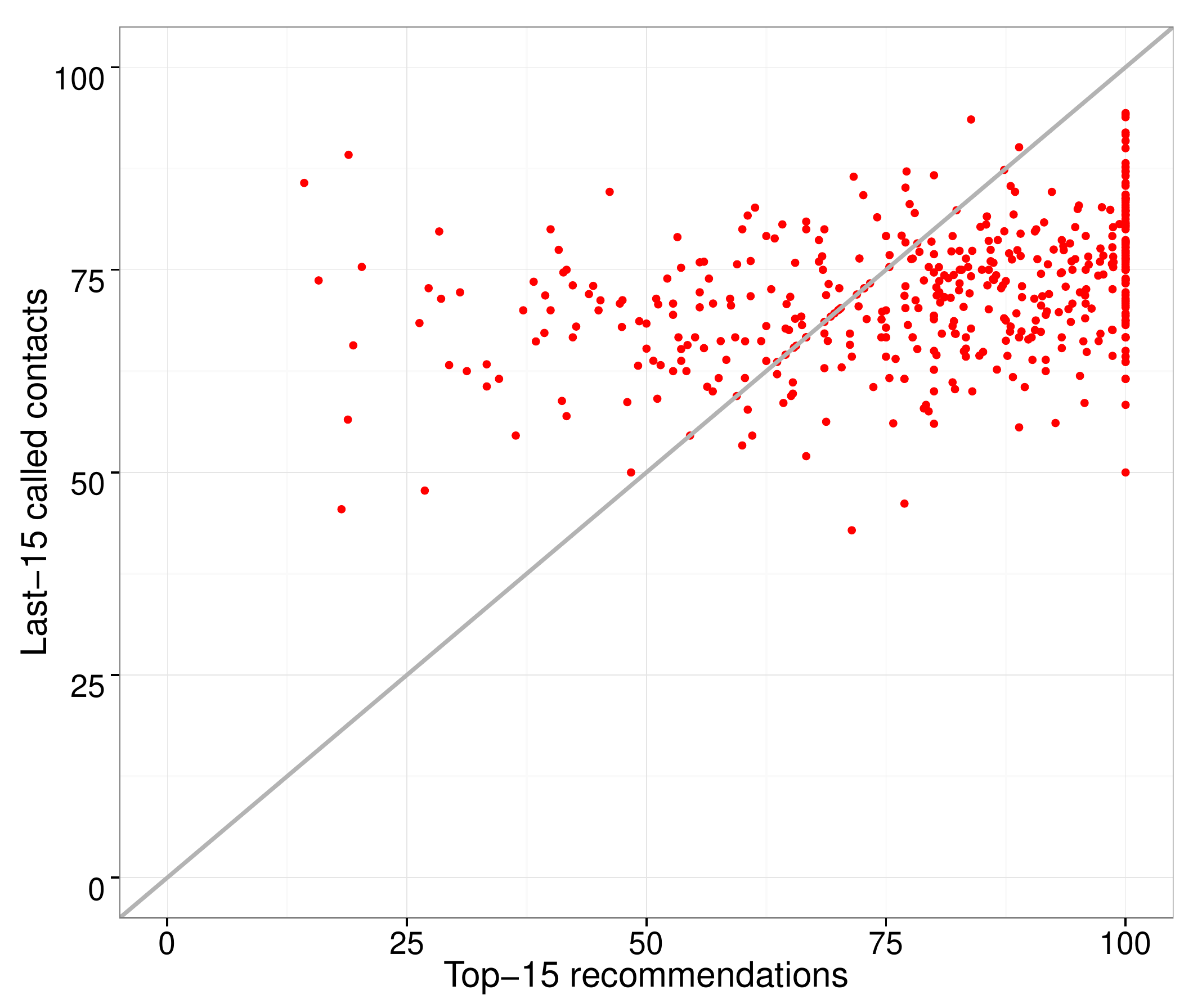}\label{subfig:top15sp2}}\hspace{1em}
 \caption{These plots show accuracy of: top-$k$ recommendations against top-$k$ called numbers and last-$k$ numbers for each user. Points below the identity line indicate that top-$k$ recommendations has better performance against the respective baseline method. Performance is reported for: 
 \protect\subref{subfig:top5rm1}, \protect\subref{subfig:top5rm2} Reality Mining - $k=5$.
   \protect\subref{subfig:top10rm1}, \protect\subref{subfig:top10rm2} Reality Mining - $k=10$.
  \protect\subref{subfig:top15rm1}, \protect\subref{subfig:top15rm2} Reality Mining - $k=15$.
 \protect\subref{subfig:top5sp1}, \protect\subref{subfig:top5sp2} Smartphone - $k=5$.
 \protect\subref{subfig:top10sp1}, \protect\subref{subfig:top10sp2} Smartphone - $k=10$.
  \protect\subref{subfig:top15sp1}, \protect\subref{subfig:top15sp2} Smartphone - $k=15$.
   }
  \label{fig:top6figs}
 \end{figure*}

 \begin{table*}[t]
 
   \caption{ Average accuracy (\%) with different methods.}
    \label{tab:accuracycomparison}
  \centering
 
  \begin{tabular}{p{3.5cm}|p{0.9cm}|p{0.9cm}|p{0.9cm}|p{0.9cm}|p{0.9cm}|p{0.9cm}}
  
   \cline{1-7}
    \multicolumn{1}{l|}{} & \multicolumn{3}{c|}{Reality Mining} & \multicolumn{3}{c}{Smartphone} \\ \cline{2-7}
    & $k=5$ & $k=10$ & $k=15$ & $k=5$  & $k=10$ & $k=15$  \\
  \hline \hline
   
   Top-$k$ called numbers & 60.04 & 70.20 & 77.65 & 45.65 & 63.73 & 74.51  \\
   \hline
  Last-$k$ numbers  &63.78 & 69.59 & 73.58& 63.94 & 69.76  & 72.83  \\
     
  \hline
  Top-$k$ recommendations & 78.70& 85.66 & 88.46 & 63.96 & 74.72  & 82.08  \\ 
     \hline
  
  \end{tabular}
 
  \end{table*}

\section{Performance Analysis}
\label{sec:analysis}

Predictions are made for the users who have at least $50$~communication events in the dataset. Hence we analyzed $89$ users in the Reality Mining Dataset and $604$ users in the Smartphone dataset. Further, we want to improve accuracy using fewer dimensions. For the last calls related features, we have used data pertinent to only the last two calls since there is a trade-off between adding dimensions to the feature set and efficiency.  

\subsection{ Top-$k$ recommendations}

From the users' perspective, top-$k$ recommendations should be more accurate as compared to last-$k$ calls. 
We generate a list of most likely numbers to be dialed at any given time: the`top-$k$ recommendations'. We compare the accuracy of top-$k$ recommendations with the accuracy obtained by last-$k$ calls. We show the performance of our approach for individual users for varying list lengths(5, 10 and 15). In Figure \ref{fig:top6figs}, x-axis represents the users (egos) in the dataset. For every user in the datasets, we report the accuracy achieved by top-$k$ recommendations vs. last-$k$ calls and top-$k$ called numbers (most frequently called contacts). The accuracy is reported for each user in \textit{Reality Mining dataset:} points in blue; and \textit{Smartphone dataset:} points in red. A higher concentration of points below the identity line indicates that top-$k$ recommendations has better performance against the respective method. Table \ref{tab:accuracycomparison} reports the average performance of our approach along with performance achieved by baseline methods, whereas, Table \ref{tab:k=10} reports the proportion of correctly predicted calls for various list lengths.

\subsection{Prediction deviation}

From the service providers' perspective, accurate prediction of calls would enable them to predict users' behavior and predict periods of high usage which in turn would lead to better load balancing and, hence, better service quality. Table \ref{tab:predictionaccuracy} reports the prediction accuracy for given deviation thresholds. These results show that a reasonable proportion of the phone calls are predictable using the proposed method. For the Reality Mining dataset 44\% of the outgoing calls were predicted below one hour error threshold. For the Smart phone dataset 31\% of the outgoing calls were predicted below one hour error threshold.



 \begin{table}[h]
 \caption{Proportion of correctly predicted calls for various list lengths for $\epsilon(1Hr.)$.}
 \centering
 \begin{tabular}{  c | c | c}
 \hline
 
         $ k$ &Reality Mining & Smartphone \\
 \hline
 \hline
   1 &  0.44   & 0.31  \\
 \hline
      2 & 0.64 & 0.49  \\
 \hline
   3 &  0.75   & 0.55  \\
 \hline
      4 & 0.77 & 0.59  \\
 \hline
   5 &  0.78   & 0.63  \\
 \hline
      6 & 0.80 & 0.66  \\
 \hline
   7 &  0.81   & 0.69  \\
 \hline
      8 & 0.83 & 0.71  \\
 \hline
   9 &  0.84   & 0.73  \\
 \hline
   10 & 0.85 & 0.74  \\
 \hline
 \end{tabular}
 
 \label{tab:k=10}
 \end{table}

\section{Discussion}
\label{sec:discussion}
Mobile phones represent one of the most commonly used communication medium. The portable nature of the medium means very little can be assumed about the situation in which the phone is used; a typical user makes calls in all kinds of
contexts. These two factors, frequency and versatility of use,
necessitate an extremely efficient call-making interface design.
 As a first step we analyzed using a machine learning approach and using few dimensions whether it is possible to predict the calling behavior of mobile phone users (given the time based features). We have identified the day of the week and time as two important features which help in accurately predicting the next outgoing call. This is supported by the fact that human interaction behavior follows a circadian rhythm. We have also analyzed the situations where it is more probable that the user calls a number from one of the last called numbers.

Evaluation methods similar to ours have been used in previous studies. With a few exceptions, most previous studies used different datasets for analyzing calling behavior, therefore, a direct comparison is not equitable. Phithakkitnukoon et al. \cite{phithakkitnukoon2011behavior} predicted the outgoing and incoming calls on Reality Mining dataset. Out of the $94$ users, they selected only $30$ users for experiments. The identities of those users is not disclosed in the paper, therefore, a direct comparison with their results is not possible. For completion, we have reported the performance of our method on $89$ out of $94$ users. The remaining $5$ users had less than $50$ communication events. For outgoing call prediction, they also generated a list of most likely numbers to be dialed at any given time. For the $30$ random users in their experiments they achieved an accuracy of $41\%$ if the predicted list is only allowed one entry. If the predicted list has five entries their model correctly predicted the dialed number $70 \%$ of the time. On the Reality Mining dataset we achieve an accuracy of $44\%$ when the top-$k$ list has one entry. Our results show more than $78\%$ accuracy on the Reality Mining dataset when the predicted list is allowed $5$ entries. Figure \ref{fig:top6figs} and Table \ref{tab:accuracycomparison} shows that our approach also performs better than the last-$k$ calls on both the datasets. 

Barzaiq et al. \cite{barzaiq2011adapting} modeled the historic call patterns of users and achieved a $35\%$ accuracy for call prediction. Haddad et al. \cite{ Had14}, reports the prediction accuracy for certain time-deviation thresholds on a dataset consisting of more than seven thousand users. Their model predicted about $17\%$ of the outgoing calls with an error below one hour. The results for our approach, reported in Table \ref{tab:predictionaccuracy}, show $44\%$ and $31\%$ accuracy for Reality Mining and Smartphone datasets respectively. Haddad et al. assumes that the call arrival patterns have a Poisson distribution. An initial analysis in \cite{kim2013analyzing} on another large dataset suggest that the call arrival process is not Poisson for all caller-callee pairs\cite{doran2015propagation}. It can be argued that the particular pattern Haddad et al. mentioned in the paper was an artifact of their dataset. 

In the previous models such as the ones proposed in \cite{Had14} and \cite{phithakkitnukoon2011behavior}, a baseline comparison was missing. The motivation behind our study was to come up with a method that could better predict the next call. Hence, from the user's point of view we found it imperative to check the performance of the last-$k$ calls as well. It is a reasonable expectation that a call prediction approach should perform better than the current approach used for smartphone call logs i.e., displaying the recent calls in chronological order. Table \ref{tab:accuracycomparison} shows that our approach performs better than the last-$k$ calls and most frequently called numbers list.

\section{Conclusions and Future Work}
\label{sec:conclusion}

We proposed a machine learning approach which uses temporal as well as last-calls features to predict the future outgoing calls. We tested our approach on two datasets from two countries and found that majority of outgoing phone calls can be predicted based on the temporal calling patterns. Our approach outperformed the two baseline approaches i.e. predicting next call based on last-$k$ calls and predicting next call using the most frequently called numbers' list. We found it very intriguing  as it opens many exciting research questions. One of them is to see whether these results can be replicated if we take a large representative sample that can be generalized to all mobile phone users. In order to deeply understand the phone call behaviour, we are in the process of collecting a large call logs dataset along with other relevant information such as demographic, geographical, and socio-economic data.  Another future research possibility could be an attempt to redesign the calling interface for mobile phones which could improve the user experience significantly. Such an interface, theoretically, would know the most likely people one is going to call at a given time and day.  In future we would like to study how users respond to an improved call log interface. A usability study in this regard is underway.



\bibliographystyle{abbrv}
%


%

\bibliography{patternbib}

\begin{thebibliography}{10}

\bibitem{ourpaper}
Anonymous, details omitted due to double-blind reviewing.

\bibitem{aledavood15b}
T.~Aledavood, E.~L{\'o}pez, S.~G.~B. Roberts, F.~Reed-Tsochas, E.~Moro,
  R.~I.~M. Dunbar, and J.~Saram{\"a}ki.
\newblock Channel-specific daily patterns in mobile phone communication.
\newblock {\em ArXiv e-prints}, July 2015.

\bibitem{aledavood15a}
T.~Aledavood, E.~{L{\'o}pez}, S.~G.~B. Roberts, F.~Reed-Tsochas, E.~Moro,
  R.~I.~M. Dunbar, and J.~Saram{\"a}ki.
\newblock Daily rhythms in mobile telephone communication.
\newblock {\em ArXiv e-prints}, Feb. 2015.

\bibitem{barabasi05}
A.-L. Barabasi.
\newblock The origin of bursts and heavy tails in human dynamics.
\newblock {\em Nature}, 435(7039):207--211, 2005.

\bibitem{barzaiq2011adapting}
O.~O. Barzaiq and S.~W. Loke.
\newblock Adapting the mobile phone for task efficiency: the case of predicting
  outgoing calls using frequency and regularity of historical calls.
\newblock {\em Personal and Ubiquitous Computing}, 15(8):857--870, 2011.

\bibitem{bergman2012you}
O.~Bergman, A.~Komninos, D.~Liarokapis, and J.~Clarke.
\newblock You never call: Demoting unused contacts on mobile phones using dmtr.
\newblock {\em Personal and Ubiquitous Computing}, 16(6):757--766, 2012.

\bibitem{calladdressbook}
A.~Bhamidipaty and P.~Deepak.
\newblock Symab: symbol-based address book for the semi-literate mobile user.
\newblock In {\em Human-Computer Interaction--INTERACT 2007}, pages 389--392.
  Springer, 2007.

\bibitem{addressbookbad}
M.~B{\"o}cker and A.~Suwita.
\newblock Evaluating the siemens c10 mobile phone: going beyond “quick and
  dirty” usability testing.
\newblock {\em Proceedings of HFT’99}, 1999.

\bibitem{candia2008uncovering}
J.~Candia, M.~C. Gonz{\'a}lez, P.~Wang, T.~Schoenharl, G.~Madey, and A.-L.
  Barab{\'a}si.
\newblock Uncovering individual and collective human dynamics from mobile phone
  records.
\newblock {\em Journal of Physics A: Mathematical and Theoretical},
  41(22):224015, 2008.

\bibitem{ctia15}
CTIA.
\newblock Annual wireless industry survey.
\newblock Technical report, CTIA - The wireless association, 2015.

\bibitem{doran2015propagation}
D.~Doran and V.~Mendiratta.
\newblock Propagation models and analysis for mobile phone data analytics.
\newblock In {\em Propagation Phenomena in Real World Networks}, pages
  257--292. Springer, 2015.

\bibitem{eagle06}
N.~Eagle and A.~Pentland.
\newblock Reality mining: sensing complex social systems.
\newblock {\em Personal and ubiquitous computing}, 10(4):255--268, 2006.

\bibitem{inferfriendship}
N.~Eagle, A.~S. Pentland, and D.~Lazer.
\newblock Inferring friendship network structure by using mobile phone data.
\newblock {\em Proceedings of the National Academy of Sciences},
  106(36):15274--15278, 2009.

\bibitem{Had14}
M.~R. Haddad, H.~B. Zghal, D.~Ziou, and H.~B. Gh{\'{e}}zala.
\newblock A predictive model for recurrent consumption behavior: An application
  on phone calls.
\newblock {\em Knowl.-Based Syst.}, 64:32--43, 2014.

\bibitem{jiang2013calling}
Z.-Q. Jiang, W.-J. Xie, M.-X. Li, B.~Podobnik, W.-X. Zhou, and H.~E. Stanley.
\newblock Calling patterns in human communication dynamics.
\newblock {\em Proceedings of the National Academy of Sciences},
  110(5):1600--1605, 2013.

\bibitem{jo12}
H.-H. Jo, M.~Karsai, J.~Kert{\'e}sz, and K.~Kaski.
\newblock Circadian pattern and burstiness in mobile phone communication.
\newblock {\em New Journal of Physics}, 14(1):013055, 2012.

\bibitem{kim2013analyzing}
H.~Kim, H.~Zang, and X.~Ma.
\newblock Analyzing and modeling temporal patterns of human contacts in
  cellular networks.
\newblock In {\em Computer Communications and Networks (ICCCN), 2013 22nd
  International Conference on}, pages 1--7. IEEE, 2013.

\bibitem{lenhart10}
A.~Lenhart.
\newblock Cell phones and american adults.
\newblock Technical report, Pew Research Center, 2010.

\bibitem{ljung1978measure}
G.~M. Ljung and G.~E. Box.
\newblock On a measure of lack of fit in time series models.
\newblock {\em Biometrika}, 65(2):297--303, 1978.

\bibitem{rsvm}
D.~Meyer, E.~Dimitriadou, K.~Hornik, A.~Weingessel, and F.~Leisch.
\newblock {\em e1071: Misc Functions of the Department of Statistics,
  Probability Theory Group (Formerly: E1071), TU Wien}, 2015.
\newblock R package version 1.6-7.

\bibitem{phithakkitnukoon2011behavior}
S.~Phithakkitnukoon, R.~Dantu, R.~Claxton, and N.~Eagle.
\newblock Behavior-based adaptive call predictor.
\newblock {\em ACM Transactions on Autonomous and Adaptive Systems (TAAS)},
  6(3):21, 2011.

\bibitem{urbanmobility}
A.~Sevtsuk and C.~Ratti.
\newblock Does urban mobility have a daily routine? learning from the aggregate
  data of mobile networks.
\newblock {\em Journal of Urban Technology}, 17(1):41--60, 2010.

\bibitem{wu10}
Y.~Wu, C.~Zhou, J.~Xiao, J.~Kurths, and H.~J. Schellnhuber.
\newblock Evidence for a bimodal distribution in human communication.
\newblock {\em Proceedings of the national academy of sciences},
  107(44):18803--18808, 2010.

\bibitem{zeru85}
E.~Zerubavel.
\newblock {\em Hidden Rhythms: schedules and calendars in social life}.
\newblock Univ of California Press, 1985.

\bibitem{zhang13}
Z.~Zhang, H.~Lin, K.~Liu, D.~Wu, G.~Zhang, and J.~Lu.
\newblock A hybrid fuzzy-based personalized recommender system for telecom
  products/services.
\newblock {\em Information Sciences}, 235:117--129, 2013.

\end{thebibliography}


\balance

\end{document}